# Catalytically active NaNbO$_3$ nanorods for sonodynamic cancer therapy

Xinyi Ding[a], Meiqi Chang[b,*]

Department of Materials Science, Fudan University, Shanghai 200433, P. R. China.

Laboratory Center, Shanghai Municipal Hospital of Traditional Chinese Medicine, Shanghai University of Traditional Chinese Medicine, Shanghai 200071, P. R. China.
Email: changmeiqi@vip.sina.com.

**Abstract**

Sonodynamic therapy (SDT) has received a lot of interest due to its deep tissue penetration and lack of invasiveness. However, SDT still prioritizes the creation of highly effective, multifunctional, and biocompatible sonosensitizers to improve the therapeutic efficiency. In this study, sodium niobate ($NaNbO_3$) nanosonosensitizers are rationed synthesized for SDT for the first time. $NaNbO_3$ nanosonosensitizers with semiconductor characteristics are proved to generate large amounts of reactive oxygen species and induce cell apoptosis under ultrasound irradiation. *In vitro* anti-tumor theranostic results confirm the mitochondrial dysfunction-dependent death pathway. *In vivo* tumor xenograft evaluation demonstrates that $NaNbO_3$ will massively induce cytotoxicity and tumor eradication under ultrasound irradiation. These results provide the paradigm of the utilization of novel nanosonosensitizers as a therapeutic nanoplatform in treating breast cancer cells.

**Introduction**

Cancer undoubtedly remains one of the most serious global health challenges in the past few decades.[1] The external minimally invasive or non-invasive treatment methods have emerged as an attractive trend due to their well-known therapeutic benefits.[2] Sonodynamic therapy (SDT) as a promising approach to cancer therapy, has generated a great deal of interest due to its low cost, safety, tissue penetration depth.[3] It stimulates sonosensitizers to release fatal reactive oxygen species (ROS) that kill tumor cells using ultrasound (US) as the exogenous energy.[4] However, the lack of suitable sonosensitizers has kept SDT in its infancy and limited its treatment efficiency.[5, 6] Consequently, millions efforts have been made to develop the proper sonosensitizers possessing both biological security and strong ROS generation productivity. There are two primary categories for the reported sonosensitizers: organic and inorganic sonosensitizers.[7] Distinguished from organic sonosensitizers with limited water solubility, poor stability, and fast body clearance,[8] inorganic sonosensitizers are promising due to the unique physiochemical properties and reduced phototoxicity, which can be identified as the outstanding SDT candidates.[9]

Particularly, the goal of researchers is to create and develop various types of semiconductor nanosonosensitizers. For instance, ultrafine titanium monoxide nanorods were rationally fabricated by Chen *et al.* for enhanced SDT.[8] Moreover, W-doping narrows the band gap and introduces oxygen and Ti vacancies of $TiO_2$ nanorods, enlarging SDT effect.[10] Shim *et al.* prepared 2D $WS_2$ nanosheets for mitochondria-targeted US-related cancer therapy.[11] Similarly, $NaNbO_3$ as a typical material with both semiconductor and pyroelectric properties is an excellent candidate with excellent optical, ionic conductivity, and photorefractive properties.[12-15] In previous work, $NaNbO_3$ semiconductors have been applied in photocatalysis, photoelectrochemical water splitting and antibacterial applications.[16-18] However, to our best knowledge, $NaNbO_3$ as a nanosonosensitizers for sonodynamic tumor therapy have been rarely reported.

Herein, we discovered NNO nanorods *via* a simple hydrothermal method for sonodynamic cancer treatment for the first time. The ROS generation ability of NNO under US irradiation was investigated by the 1,3-Diphenylisobenzofuran **(**DPBF) degradation and electron spin resonance (ESR) experiments, indicating the production of $^1O_2$ and •OH. Meanwhile, the killing effect of

SDT with NNO nanomaterials at cellular level was examined through CCK-8 assays, intracellular ROS production measurement and live/dead staining experiments. Moreover, *in vitro* anti-tumor theranostic results confirm the mitochondrial dysfunction-dependent apoptosis pathway (**Scheme 1**). Finally, *in vivo* SDT with NNO provides highly promising potential for tumor eradication. To summarize, this paradigm might motivate designers to create more versatile semiconductor sonosensitizers for theranostics against serious illnesses.

**Experimental section**

**Synthesis of NNO nanomaterials.** Firstly, 12 M of sodium hydroxide aqueous solution (40 mL) and 60 mM of niobium pentaoxide aqueous solution (40 mL) were ultrasonically mixed for 2 h. The mixture was then transferred into an autoclave and treated at 160 ℃ for 4 h. Then, the obtained products were washed with deionized water and ethanol, and dried at 80 ℃. The final products were calcined at 700 ℃ for 6 h.

**Characterization.** Transmission electron microscopic (TEM) photographs were conducted on a JEM-2100Felectron microscope (200 kV). Energy dispersive X-ray spectroscopy (EDS) and relevant element mapping analysis were obtained by a JEOL ARM-300F with spherical aberration correction operated at 300 kV. X-ray diffraction (XRD) was measured on a Rigaku D/MAX-2200 PC XRD system (parameters: Cu K$\alpha$, $\lambda$=1.54 Å, 40 mA, 40 kV). X-ray photoelectron spectroscopy (XPS) spectrum was obtained on an ESCAlab250 (Thermal Scientific, US). Ultraviolet-visible light-near-infrared (UV-vis-NIR) absorbance spectra were determined by a UV-3600 Shimadzu UV-vis-NIR spectrometer. Confocal laser scanning microscopy (CLSM) images were scored on a FV1000 (Olympus Company, Japan). Cell apoptosis assays were captured by a BD LSRFortessa flow cytometry (BD, US).

**Sonodynamic effect of NNO nanomaterials on the degradation of 1,3-Diphenylisobenzofuran (DPBF).** Initially, NNO aqueous dispersion (60 ppm) and DPBF ethanol solution (1000 ppm) were prepared. The whole sonodynamic experiment was carried out in the dark to avoid the effect of photolysis. 3 mL of NNO aqueous dispersion and 260 μL of DPBF ethanol solution were mixed well. The well-dispersed mixture was analyzed by UV-vis absorption spectroscopy. The degradation rate can be calculated as:

$$Degradation\ rate = (A_{0min} - A_{nmin})/A_{0min} \times 100\%$$

$A_{0min}$ and $A_{nmin}$ are the absorption value at 420 nm before and after US irradiation.

**ESR measurement.** 100 μL NNO (10 mg mL$^{-1}$) and 20 μL DMPO (Dojindo-D048, China Co., Ltd.) were added into 900 μL H$_2$O under US irradiation. Then •OH with the characteristic 1:2:2:1 signal was detected by an ESR spectrometer.

**Cell culture and *in vitro* cytotoxicity assay.** 4T1 murine breast cancer cells, purchased from Shanghai Institute of Cells, Chinese Academy of Sciences, were cultured in Roswell Park Memorial Institute 1640 Medium (RPMI 1640, Gibco) containing 10% fetal bovine serum (FBS, Gibco) and 1% penicillin-streptomycin in a humidified incubator (5% CO$_2$, 37 ℃). 4T1 cells were generally seeded in a cell culture flask (Corning, USA) for 24 h. Varied concentrations of NNO (0, 6.25, 12.5, 25, 50, 100, 200, 400, 800 μg mL$^{-1}$, n = 9) were co-incubated with 4T1 breast cancer cells for 24 h. After treating with different groups, a standard CCK-8 viability assay (Cell counting Kit-8, Beyotime) was conducted to evaluate the *in vitro* cytotoxicity of NNO.

**SDT effect at intracellular level.** 4T1 cells were seeded into 96-well plates (cell density = 10$^4$ cells per disk) and cultured for 24 h. Then, 4T1 cells were treated with the following conditions: 1. Control; 2. US (1.0 MHz, 1.5 W·cm$^{-2}$, 50% duty cycle, 3 min); 3. NNO (100 μg mL$^{-1}$); 4. NNO + US. After co-incubation at 37 °C for 5 h, 4T1 cells cultured with NNO were exposed to US irradiation. After another 12 h, the cell viability of treated cells was quantified by CCK-8 assay.

**Detection of live/dead cells.** 4T1 cells were seeded in a confocal glass bottom-dish and cultured for 24 h. Then, 4T1 cells were treated with the following conditions: 1. Control; 2. US (1.0 MHz, 1.5 W·cm$^{-2}$, 50% duty cycle, 3 min); 3. NNO (100 μg mL$^{-1}$); 4. NNO + US. After co-incubation at 37 °C for 12 h, the cells were co-incubated with Calcein AM and PI for 30 min (Dojindo Molecular Technologies) followed by observation with CLSM.

**Detection of intracellular ROS production.** 4T1 cells were seeded in a confocal glass bottom-dish and cultured for 24 h. Then, 4T1 cells were treated with the following conditions: 1. Control; 2. US (1.0 MHz, 1.5 W·cm$^{-2}$, 50% duty cycle, 3 min); 3. NNO (100 μg mL$^{-1}$); 4. NNO + US. After co-incubation, DCFH-DA (Beyotime Biotechnology, 100 μL, 10% in the culture medium) was added followed by an additional incubation. Finally, the cells were washed with PBS two times and were observed by CLSM.

**Cell apoptosis analysis.** 4T1 cells were seeded in 6-well plates and cultured for 24 h. Then, 4T1 cells were treated with the following conditions: 1. Control; 2. US (1.0 MHz, 1.5 W·cm$^{-2}$, 50% duty cycle, 3 min); 3. NNO (100 μg mL$^{-1}$); 4. NNO + US. Then, 4T1 cells were rinsed gently by PBS for 3 times to wipe off NNO remnants and double stained with the Annexin V-FITC Apoptosis Detection Kit (Beyotime-C1062S, Shanghai, China) followed by the flow cytometric analysis.

**Detection of mitochondrial membrane potential.** 4T1 cells were seeded in a confocal glass bottom-dish and cultured for 24 h. Then, 4T1 cells were treated with the following conditions: 1. Control; 2. US (1.0 MHz, 1.5 W·cm$^{-2}$, 50% duty cycle, 3 min); 3. NNO (100 μg mL$^{-1}$); 4. NNO + US. After different treatments, 4T1 cells were stained with the Mitochondrial membrane potential assay kit (Beyotime-C2006, Shanghai, China) with JC-1 followed by CLSM.

***In vivo* evaluation of SDT efficiency with NNO.** The animal experiment was conducted with the approval of ethics by Ethic Committee of Shanghai University. Firstly, the tumor model was established by subcutaneously injecting female BALB/c nude mice with 4T1 breast cancer cells. When tumor sizes are about ~400 mm$^3$, these mice were divided into four groups (non-treated cells as control, US, NNO, NNO + US). Then, the mice in each group were intratumorally administered with 1000 μL dispersion at the NNO dose of 2000 μg·mL$^{-1}$ except for the control group and the US group. In US irradiated groups, the mice were treated by US irradiation (1.0 MHz, 1.5 W·cm$^{-2}$, 50 % duty cycle, 3 min) after the injection. Tumors were collected and sliced for further staining by hematoxylin and eosin (H&E) to observe the structure and status of cells, terminal deoxynucleotidyl transferase dUTP nick-end labeling (TUNEL) to detect DNA fragmentation during apoptosis, and Ki-67 antibody staining to determine the growth fraction of cells 12 h after different treatments (n = 3). The tumor volume was measured every two days based on a standard protocol (V= (ab$^2$)/2, where a and b refer to the largest length and width of tumor, respectively), with the photography of tumor and body weight also recorded accordingly.

**Results and discussion**

NaNbO$_3$ (NNO) nanocrystals were synthesized *via* a facile hydrothermal approach with subsequent calcination process using Nb$_2$O$_5$ and NaOH as raw materials (**Figure 1a**). Transmission electron microscopy (TEM) image and the enlarged section show the rod morphology of NNO (**Figure 1b**). XRD pattern of NaNbO$_3$ was measured to further confirm the orthorhombic crystalline structure (JCPDS 73-0803) (**Figure 1c**). The characteristic diffraction peaks at 22.5°, 32.3°, 46.2°, 52.2°, 57.7°, 67.9° and 72.6° correspond to the (110), (114), (220), (224), (028), (228) and (331) planes of orthorhombic NNO, respectively. EDS spectrum of NNO with distinct characteristic peaks of Na, Nb and O can be detected by **Figure 1d**. Elemental mapping (**Figure 1e**) shows uniform dispersion of Na, Nb and O elements in the nanorod structure, confirming the the successful synthesis of the final product.

XPS measurement was performed to further analyze the chemical state of NNO (**Figure 2**). The survey spectrum of NNO demonstrated the presence of Na, Nb and O (**Figure 2a**). In the high-resolution spectrum of Na, only one main peak located at 1071.18 eV has been detected, which is assigned to Na 1s (**Figure 2b**). Nd 3d spectrum contains two main peaks at 206.41 eV and 209.15 eV, which can be attributed to the characteristic binding energies of Nb $3d_{5/2}$ and Nb $3d_{3/2}$ (**Figure 2c**).[19] It should be noted that the main peak of O 1s can be deconvolved into two peaks at 529.44 eV and 531.45 eV, due to the coexistence of two forms of O, lattice oxygen ($O^{2-}$) and surface oxygen, respectively (**Figure 2d**).[20]

UV-vis spectrum was conducted to study the optical and semiconductor characteristics of NNO (**Figure 3a**). The synthesized products exhibited a broad optical absorption below 400 nm. Tauc's plot, as the most common and accurate method of calculating the bandgap of semiconductors, is used through the UV-vis absorption spectrum.[21] This method is realized by extrapolating the linear fit of (αhν)r versus hν, in which α is the absorption coefficient. The parameter r is 0.5 or 2, which depends on the indirect or direct features of semiconductor, and hν is the photon energy.[22] The synthesized orthorhombic NaNbO3 is an indirect band-gap semiconductor, so the parameter r equals 0.5. The band gap energy of NNO was calculated to be about 3.25 eV (Figure 3b). Therefore, No absorption in visible light region could be attributed to the relatively high band gap energy. Considering the role of external ultrasound, NNO with semiconductor characteristics could achieve the electron transition process from the valence to the conduction band, leaving holes in valence band. Then, the electron-hole pairs would migrate to the surface of NNO and react with $O_2$ and $H_2O$ to form $^1O_2$ and •OH, respectively (**Figure 3b**). To clarify the ROS generation capability induced by sonodynamic effect, 1,3-Diphenylisobenzofuran (DPBF) was firstly applied for the detection of $^1O_2$. The decrease of the absorbance at 420 nm can be observed with the increase of US irradiation time in the NNO +

US group (**Figure 3c**), while a far slighter decline was detected in the US alone group (**Figure 3d**). The degradation efficiencies are 16% and 38%, respectively (**Figure 3e**), demonstrating the significant $^1O_2$ generation of NNO. The electron spin resonance (ESR) experiments were applied to research the production of •OH of NNO assisted by US irradiation (**Figure 3f**). Compared with US alone group, the characteristic signal emerged in NNO + US group, confirming the generation of •OH.

To quantitatively evaluate the *in vitro* ROS generation and therapeutical effect of NNO (**Figure 4a**), the cell-counting kit 8 (CCK-8) assays were firstly implemented to assess the in vitro cytotoxicity. NNO nanomaterials were incubated with 4T1 breast cancer cells for 24 h at pre-set concentrations, and negligible cell killing rate can be detected even at the concentration of 200 μg mL$^{-1}$ (Figure 4b). A sharp decline of cell viability can be observed at the concentration of 400 μg mL$^{-1}$, demonstrating that the cytotoxicity of NNO to 4T1 breast cancer cells cannot be ignored. Therefore, NNO groups with concentrations below 200 μg mL$^{-1}$ were applied in the following experiment under US irradiation to further investigate SDT effect. The cell viability significantly decreased to 48.2% after incubated with NNO at 100 μg mL$^{-1}$ under US irradiation (**Figure 4c**). Moreover, the relative cell viability treated with NNO + US group was dramatically declined to 29.3%, whereas the US only group and the NNO only group showed negligible decline in cell viability (**Figure 4d**). 2,7-dichlorofluorescein (DCFH-DA) probe was used to detect the generation of intracellular ROS. Negligible green fluorescence can be observed in the control group, NNO alone group and US alone group. In contrast, the obvious green fluorescence was observed in the NNO + US group, indicating ROS generation induced by SDT effect (**Figure 4e**). Furthermore, 4T1 breast cancer cells were stained with Calcein-AM (green for living cells) and PI (red for dead cells) to visualize the cell killing effect through confocal laser scanning microscopy (CLSM). Evident red fluorescence was captured in the NNO + US group, demonstrating the presence of a large number of dead cells. However, nearly no red fluorescence can be detected in the other three groups (**Figure 4f**).

A standard flow apoptosis experiment was initially used to investigate the specific antineoplastic mechanism in light of the beneficial antiproliferative effect. Distinct early apoptosis (Q3 quadrant; 53.1%) and late apoptosis (Q2 quadrant; 8.97%) signals emerged in NNO + US group, while the control group, the NNO group and the US alone group remained the high survival rates (**Figure 5a, b**), which is consistent with the results of CCK-8 measurement as well as live/dead cell staining. Furthermore, the 5,5',6,6'-tetrachloro-1,1',3,3'-tetraethyl-imidacarbocyanine iodide (JC-1) probe was used to examine the change in mitochondrial membrane potential in various groups in light of the tight association between cell death and mitochondrial dysfunction (**Figure 5c**). JC-1 probe, in which red aggregates and green monomer represent the integrated and broken mitochondrial membranes, can be used to determine whether the mitochondria are polarized. Notably, the widespread green fluorescence indicated that the US-triggered NNO nanorods extensively damaged the mitochondrial membranes (**Figure 5d**).

Finally, the *in vivo* sonodynamic therapeutic performance of NNO was investigated in 4T1 tumor-bearing mice. Four groups (control, US, NNO, NNO + US) were set to assess the SDT efficiency. US irradiation was applied after the intratumoral administration of NNO for 3 min. Consistent with the negligible cytotoxicity of NNO *in vitro*, no significant abnormal body-weight changes were found in all groups (**Figure 6a**). In the NNO + US group, the tumor growth was significantly suppressed, while tumors in the other three groups (Control group, US group, NNO group) still grew rapidly (**Figure 6b**). The extent of cell damage and necrosis was then determined by performing H&E staining experiments on tumor sections (**Figure 6c**). The typical histopathological damage and the decrease of 4T1 tumor cell nuclei emerged in the NNO + US group. Moreover, the most significant apoptosis of tumor cells was induced by the sonodynamic effect with NNO, which has been confirmed by terminal deoxynucleotidyl transferase dUTP nick-end labeling (TUNEL) experiments (**Figure 6d**). The Ki67 immunohistochemical staining assay further revealed the SDT-induced tumor proliferation inhibitory effect (**Figure 6e**).

**Conclusion**

In summary, NaNbO$_3$ nanorods were successfully synthesized by a simple hydrothermal method as a kind of new sonosensitizer for effective SDT. The outstanding ROS generation ability of NNO under ultrasound irradiation was confirmed by the DPBF degradation and ESR experiments. According to *in vitro* experiments, the CCK-8 protocol and the live/dead cell staining assay demonstrated the inhibitory effect of cancer cell proliferation. Moreover, US-mediated NNO treatment induced cell apoptosis through the mitochondrial dysfunction pathway. Finally, *in vivo* assessment of sonodynamic efficacy by NNO was completed to confirm the prominent tumor eradication under US irradiation. This work sheds light on the development of novel nanosonosensitizer for efficient cancer treatments.

**Acknowledgments**

This work was financially supported by Future Plan of Shanghai Medical Innovation and Development Foundation (Grant No. WL-ZXYJH-2022001K).

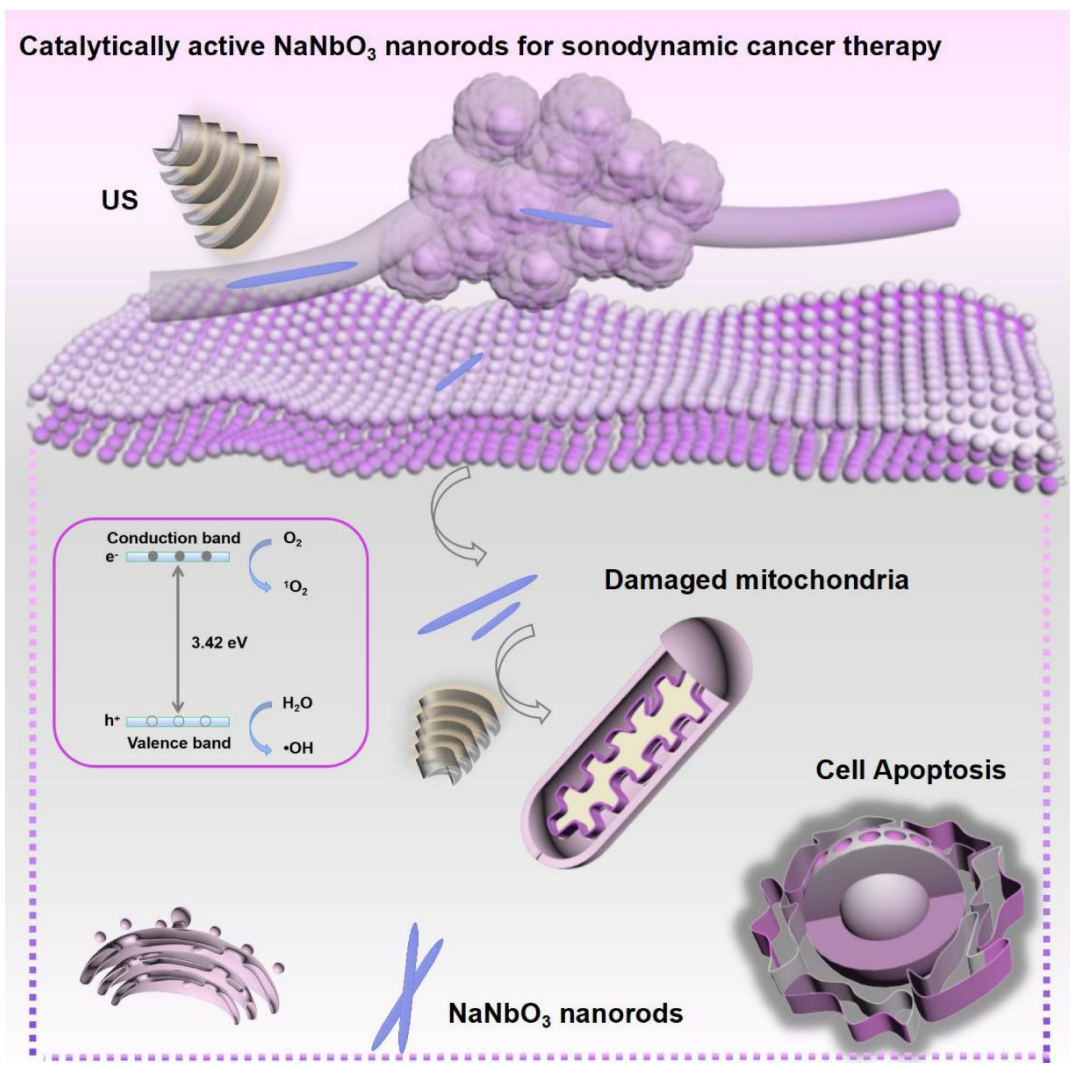

**Scheme 1.** Schematic diagram of catalytically active NaNbO$_3$ nanorods for sonodynamic cancer therapy.

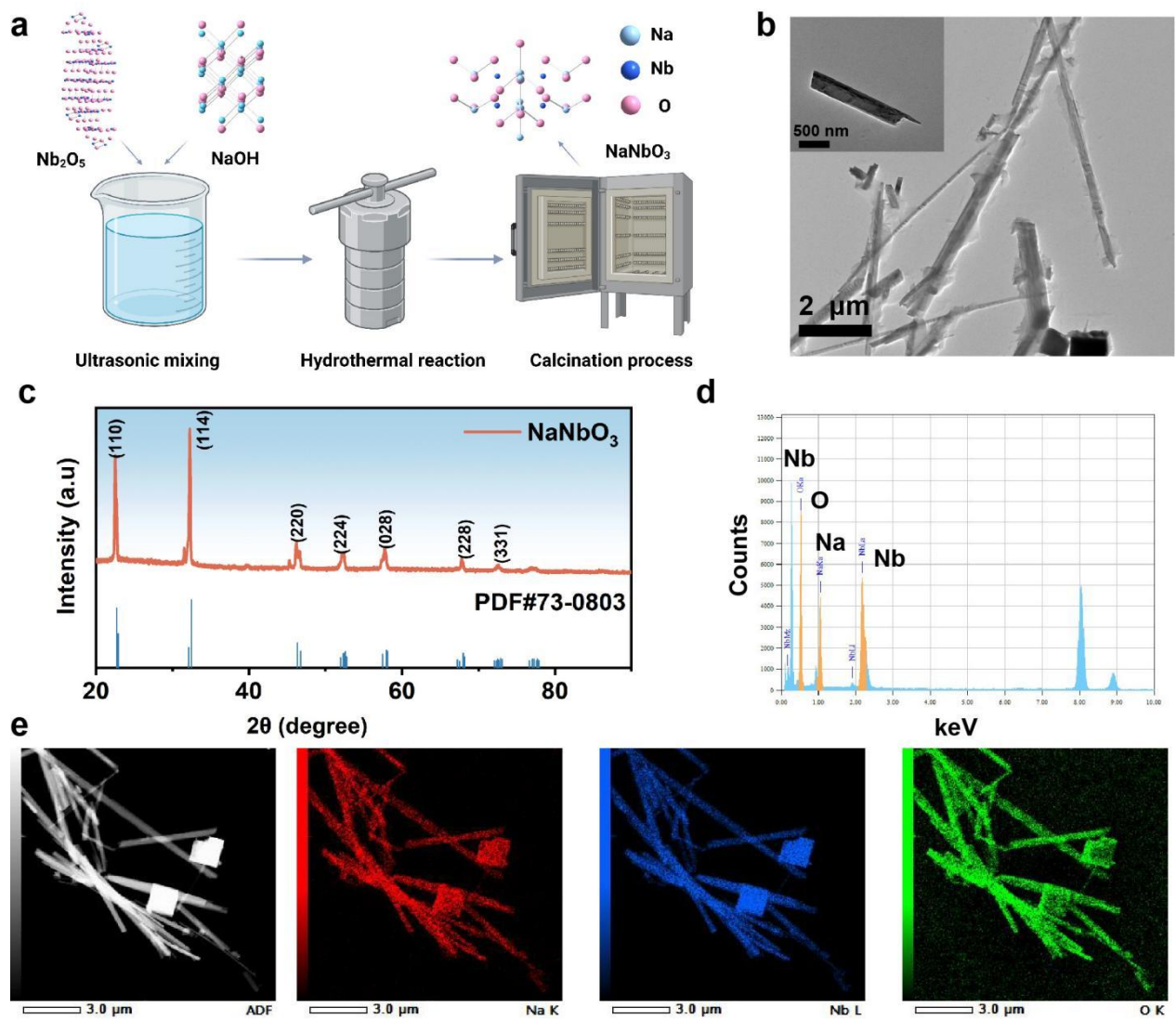

**Figure 1.** (a) Schematic illustration of the synthetic procedure of NaNbO$_3$ nanorods. (b) TEM images, (c) X-ray diffraction pattern and (d) EDS analysis of NaNbO$_3$ nanorods. (e) Elemental mapping of NaNbO$_3$.

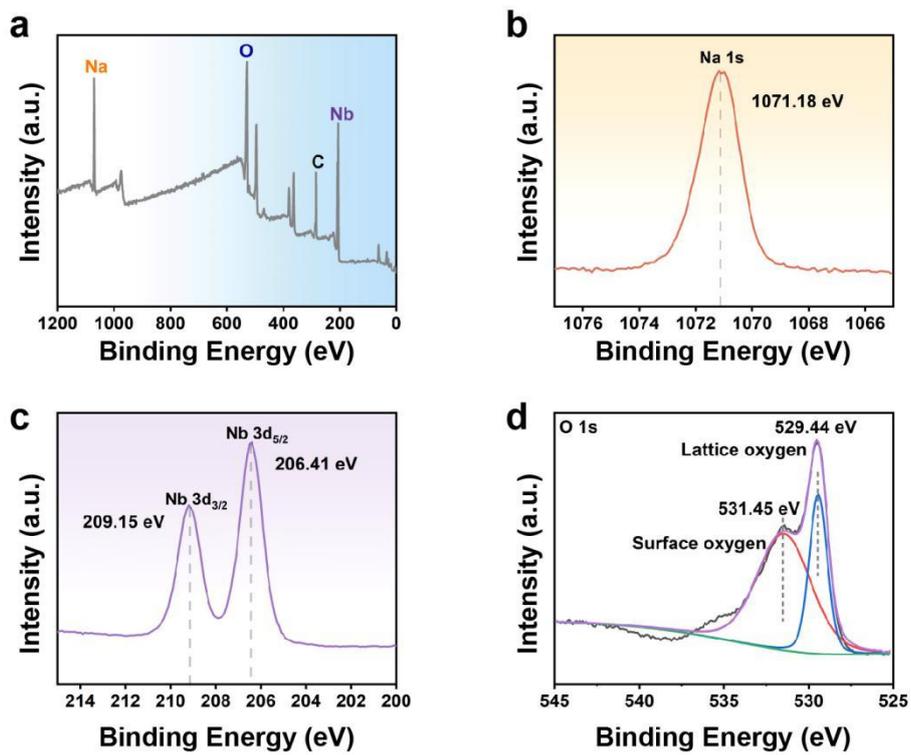

**Figure 2.** XPS spectra of (a) wide-scan, (b) Na 1s, (c) Nb 3d and (d) O 1s of NaNbO$_3$ nanorods.

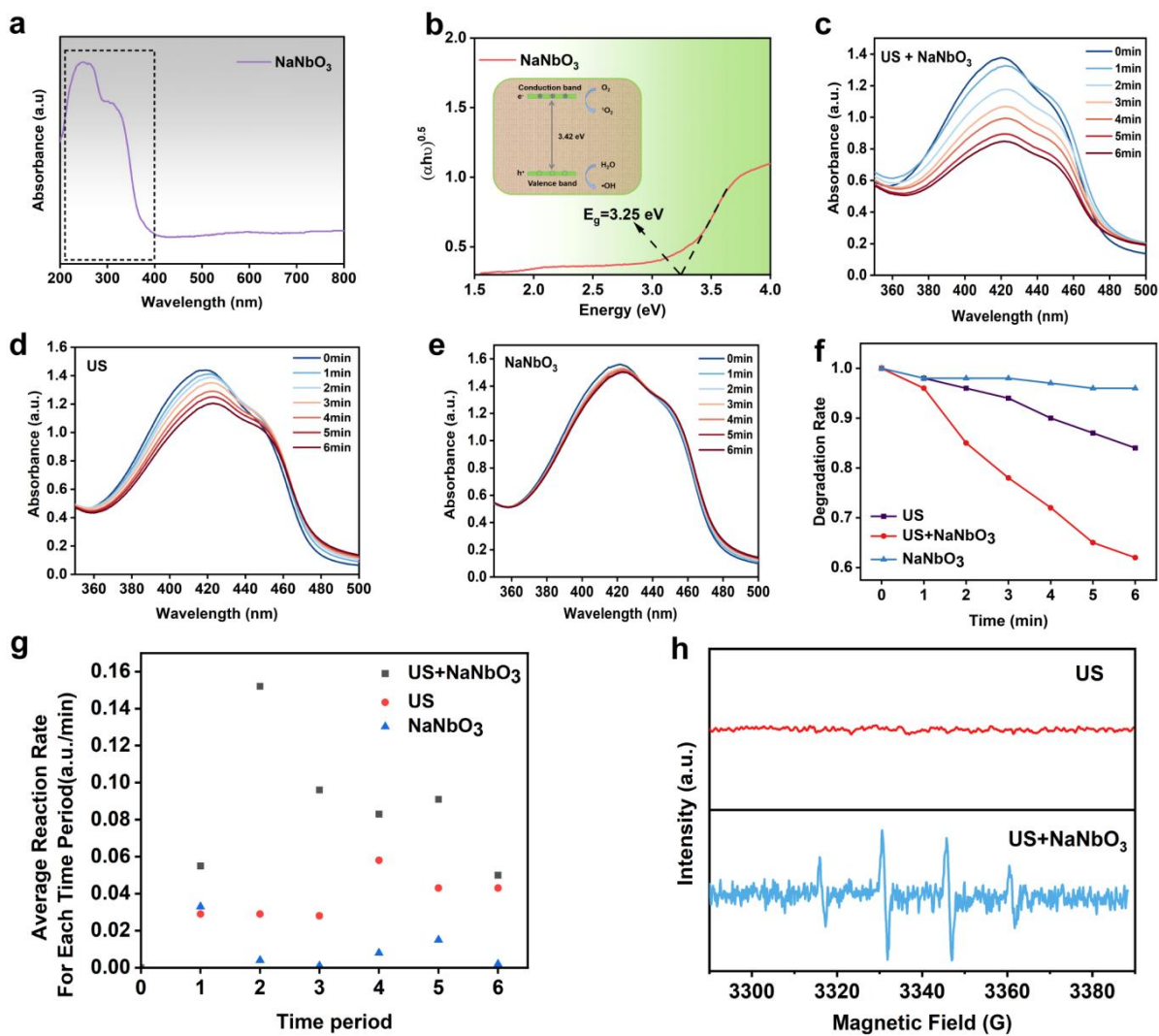

**Figure 3.** (a) UV-Vis absorbance spectrum and (b) Energy bandgap of NNO. The illustration is the proposed mechanism of ROS generation by NNO under US irradiation. DPBF fading test treated with (c) US groups, (d) US + NNO groups and (e) NNO groups. (f) Degradation of DPBF by US, US + NNO and NNO groups. (g) Average reaction rate of DPBF for each 1 min period after various treatments. (h) ESR spectra of •OH trapped by DMPO.

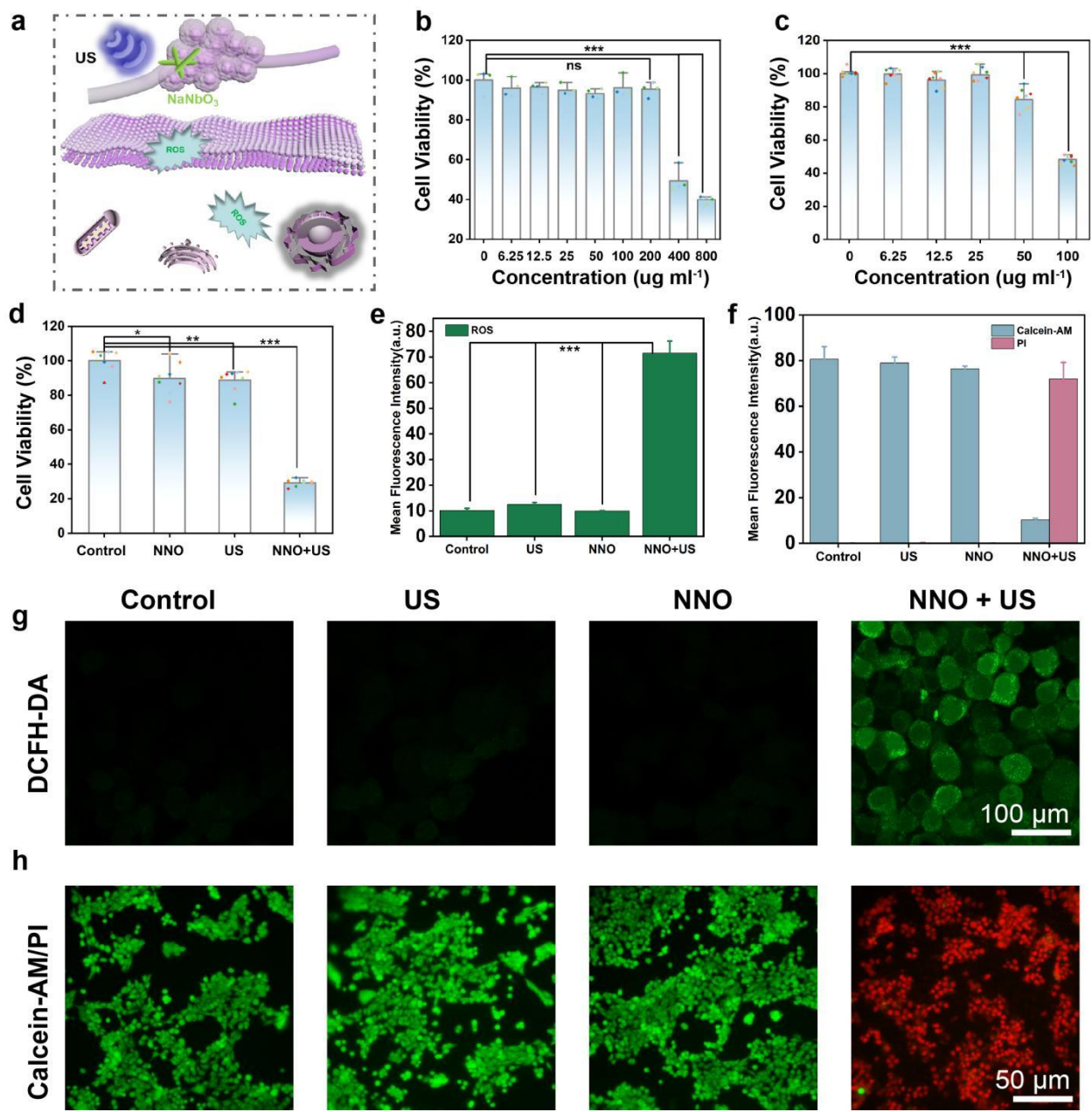

**Figure 4.** (a) Schematic illustration of *in vitro* SDT of NaNbO$_3$. (b) Cytotoxicity of NaNbO$_3$ with different concentration toward 4T1 cells. (c) Cytotoxicity of NaNbO$_3$ with different concentration toward 4T1 cells in the presence of US irradiation. (d) Relative cell viabilities of 4T1 cells after various treatments. (e) Quantification of DCFH-DA in 4T1 cells after various treatments. (f) Quantification of the mean fluorescence intensity of Calcein-AM/PI in 4T1 cells after various treatments. (g) Intracellular ROS level of 4T1 cells after different treatments. (h) Calcein AM/PI staining of 4T1 cells after different treatments. Statistical significances were calculated via Student's t-test. *p < 0.05, **p < 0.01 and ***p < 0.001.

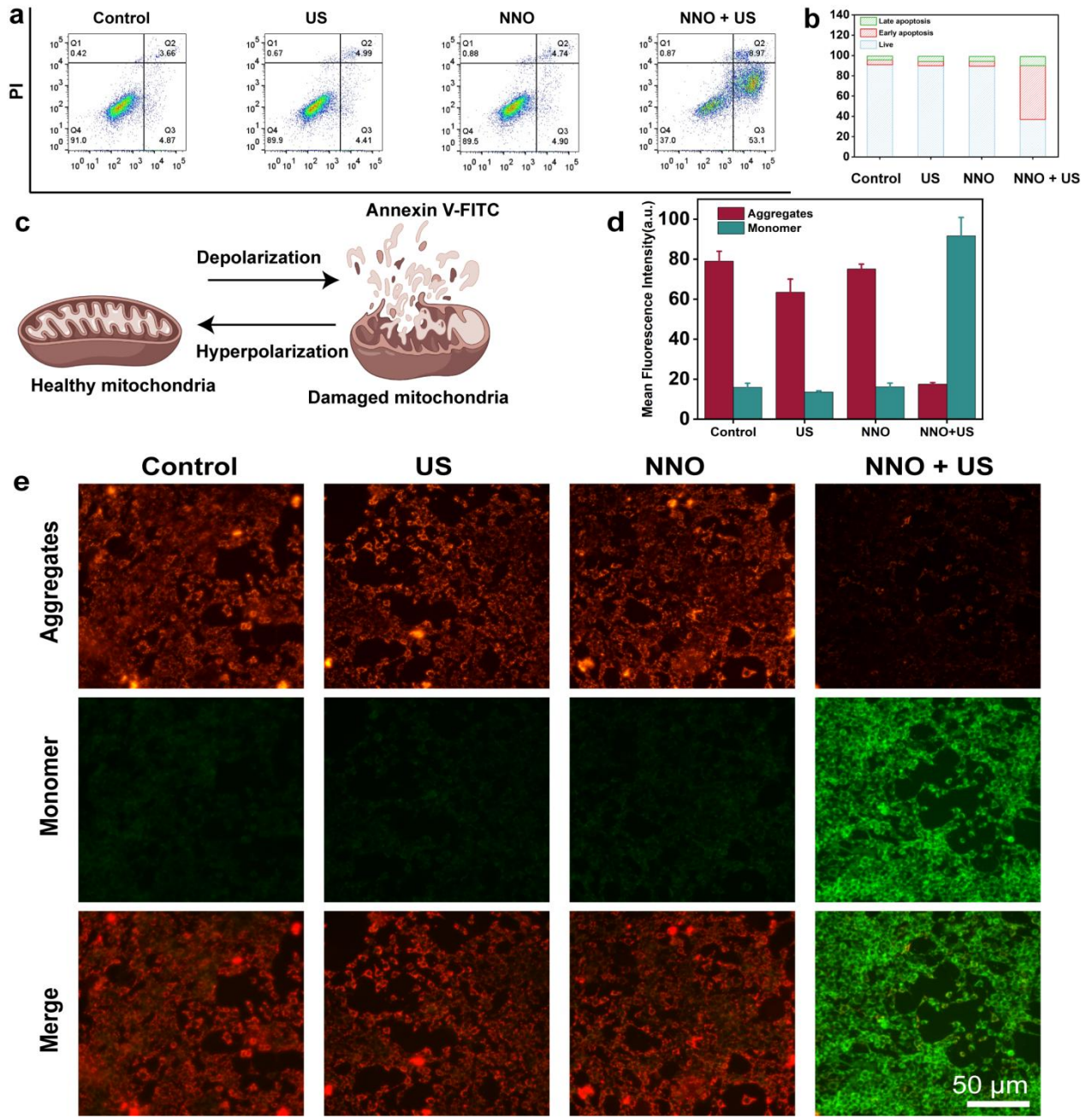

**Figure 5.** (a) Flow cytometric analysis on 4T1 cells co-stained with Annexin V-FITC and propidium iodide after different treatments. (b) The corresponding semiquantitative analysis. (c) Schematic illustration of mitochondrial membrane potential change. (d) Quantification of the mean fluorescence intensity of JC-1 staining in 4T1 cells after various treatments. (e) JC-1 staining of 4T1 cells after different treatments.

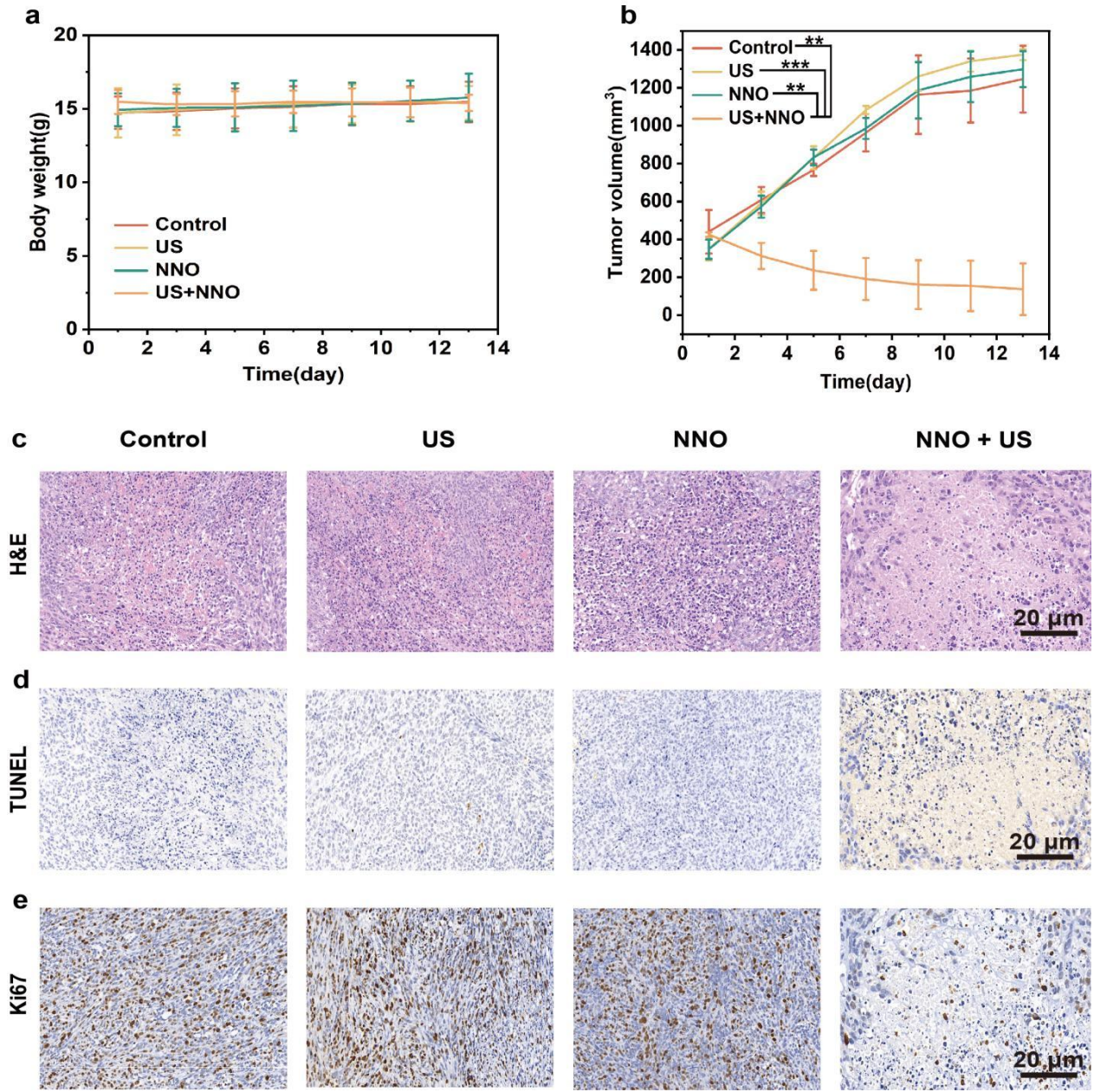

**Figure 6.** (a) Time-dependent body-weight curves. (b) Time-dependent tumor-growth curves of tumor-bearing mice after different treatments, including Control group, US group, NNO group and US+NNO group. (c) H&E staining, (d) TUNEL staining and (e) antigen Ki67 immunofluorescence staining in tumor sections. Statistical significances were calculated via Student's t-test. *$p < 0.05$, **$p < 0.01$ and ***$p < 0.001$.